\documentclass{aa}
\usepackage{txfonts}
\usepackage{graphicx}
\usepackage{afterpage}

 \newcommand{\msun}{${\rm M}_{\odot}$ }
 \newcommand{\al}{$^{26}{\rm Al}$~}
 \newcommand{\kms}{${\rm km}.{\rm s}^{-1}$~}

 \def\lesssim{\mathrel{\hbox{\rlap{\hbox{\lower4pt\hbox{$\sim$}}}\hbox{$<$}}}}
 \def\gtrsim{\mathrel{\hbox{\rlap{\hbox{\lower4pt\hbox{$\sim$}}}\hbox{$>$}}}}

\makeatletter
\let\@internalcite\cite
\def\cite{\@ifstar{\citeyear}{\citefull}}
\def\citefull{\def\astroncite##1##2{##1 ##2}\@internalcite}
\def\citeyear{\def\astroncite##1##2{##2}\@internalcite}
\def\citeau{\def\astroncite##1##2{##1}\@internalcite}
\def\citen{\def\astroncite##1##2{##1 (##2)}\@internalcite}
\def\possesivcite{\def\astroncite##1##2{##1's (##2)}\@internalcite}
\def\@citex[#1]#2{\if@filesw\immediate\write\@auxout{\string\citation{#2}}\fi
  \def\@citea{}\@cite{\@for\@citeb:=#2\do
    {\@citea\def\@citea{; }\@ifundefined
       {b@\@citeb}{{\bf ?}\@warning
       {Citation `\@citeb' on page \thepage \space undefined}}%
{\csname b@\@citeb\endcsname}}}{#1}}
\def\@cite#1#2{#1\if@tempswa , #2\fi}
\def\@biblabel#1{}
\makeatother
\authorrunning{}
\titlerunning{\al production in rotating Wolf-Rayet stars}

\begin{document}
   \title{New estimates of the contribution of Wolf--Rayet stellar winds to the 
   Galactic \al}
   \author{A. Palacios\inst{1}
     \and
     G. Meynet\inst{2}
     \and
     C. Vuissoz\inst{3}
     \and
     J. Kn\"odlseder\inst{4}
     \and
     D. Schaerer\inst{2,5}
     \and 
     M. Cervi\~no\inst{6}
     \and 
     N. Mowlavi\inst{7}
     }

   \offprints{G. Meynet or A. Palacios}

   \institute{Institut d'Astronomie et d'Astrophysique de
              l'Universit\'e Libre de Bruxelles - CP-226, Bd du
              Triomphe, 1050 Brussels (Belgium)\\
             \email{palacios@astro.ulb.ac.be}
	     \and
	     Observatoire de Gen\`eve, 51 Chemin des Maillettes, CH-1290 Sauverny (Switzerland)\\
             \email{Georges.Meynet@obs.unige.ch}
             \and 
	     Laboratoire d'Astrophysique de l'EPFL - Observatoire - CH-1290 Chavannes-des-Bois (Switzerland)
             \and
	     Centre d'\'Etude Spatiale des Rayonnements - 9, av. du Colonel
              Roche - BP 4346 - 31028 Toulouse Cedex 4 (France)
	      \and
	      Laboratoire d'Astrophysique de Toulouse-Tarbes, 14 avenue
              Edouard Belin, F-31400 Toulouse (France)  
	      \and
	      Instituto de Astrof\'{i}sica de Andaluc\'{i}a (CSIC) - Camino bajo de
              Hu\'etor 24, Apdo. 3004 - 18080 Granada (Spain)
	      \and
	      INTEGRAL Science Data Center - Chemin d'Ecogia 16- CH-1290
              Versoix (Switzerland) 
             }

   \date{Received ; accepted }


   \abstract{We present new yields of \al from Wolf--Rayet stellar winds
   based on rotating stellar models which account well for numerous
   observed properties of massive stars. We study the impacts on the yields
   of a change of initial mass, metallicity and initial rotation
   velocity. We also consider the effects of a change of mass loss rates
   during the Wolf--Rayet phase.
   
   We show that for surface rotation velocities during the core H--burning
   phase matching to the observed ones, the quantity of \al ejected by a
   star of a given initial mass and metallicity is roughly doubled when the
   effects of rotation are taken into account. The metallicity dependence
   of the yield is, on the other hand, very similar to that obtained from
   non--rotating models.
     
   We estimate that {\it at least} about 20\% to 50\% (e.g. $\sim$ 0.6 --
   1.4 M$_\odot$) of the live \al detected in the Milky--Way originates
   from Wolf--Rayet stellar winds.  We show the importance of a good
   knowledge of the present metallicity gradient and star formation rate in
   our galaxy for modeling both the variation of the \al surface density
   with the galactocentric distance and the global contribution of the
   Wolf--Rayet stellar winds to the present galactic mass of \al.
    }

   \maketitle

\section{Introduction}

The \al radionuclide, live or extinct, is observed in at least three
different types of media : 1) excesses of daughter elements from \al are
observed in presolar stardust grains (see the review by Clayton \& Nittler
\cite{CN04}); 2) evidence for injection of live \al in the early Solar
System nebula can be found in meteorites (see e.g. MacPherson et
al. \cite{Mac95}); 3) \al is also detected in the galactic interstellar
medium through its decay emission line at 1.809 MeV (Mahoney et
al. \cite{Mah82}; Diehl et al. \cite{Di95}). This diffuse emission is
observed in the plane of our Galaxy.  From its total intensity galactic \al
masses have been derived in the range from 1.5 \msun to 3 \msun, depending
much on the assumed underlying \al density distribution, and in particular,
on the yet poorly determined scale height of \al (see e.g. Fig.~4.7 in
Kn\"odlseder \cite{Kn97}).  In this paper we shall focus our attention on
the galactic interstellar \al.

Since this nuclide has a lifetime of roughly 10$^6$ yr, much shorter than
the Galaxy lifetime, it must be continuously produced by one or more
nucleosynthetic source to be observed.  The question which still remains to
be solved is what is the nature of these sources.  A correct answer to this
question is not only interesting on its own, but will enable to use the
emissivity at 1.809 MeV as a penetrating tool\footnote{ $\gamma$--rays do
not suffer any absorption by the interstellar medium.}  for studying
regions of current nucleosynthetic activity in the Milky Way.

From the observed distribution of the emission in the plane of our Galaxy
it is generally inferred that the massive stars, either through their winds
(Dearborn \& Blake \cite{DB85}; Vuissoz et al. \cite{Vu04}) or through
their ejections at the time of the supernova event (Timmes et
al. \cite{Ti95}; Woosley \& Weaver \cite{WW95}), are the main sources of
\al (see also the review by Prantzos \& Diehl \cite{PD96} and references
therein).  Here we shall only address the contribution of the winds of
Wolf--Rayet (WR) stars.\\

\noindent {\bf Wolf--Rayet versus core collapse supernovae}\\

Although WR stars end their lives as supernovae, and also contribute to the
interstellar medium enrichment of \al through explosive nucleosynthesis, we
distinguish the ``wind'' contribution from that of the ``supernova
explosion'' considering that observations allow us to disentangle the two
channels for \al production.  Indeed the ejection of \al by the winds of
massive stars is not accompanied by the ejection of $^{60}$Fe, a
radionuclide with a lifetime of 2.2 Myrs, whereas its ejection at the time
of the supernova explosion is.  Secondly, after a burst of star formation,
the enrichment of the interstellar medium by the WR stellar winds begins
prior to the enrichment by the supernovae explosions (Cervi\~no et
al. \cite{Ce00}; Pl\"uschke et al. \cite{Pl02}). This is due to the fact
that WR stars originate from more massive (and thus shorter lived)
progenitors than the bulk of the core collapse supernovae (hereafter
cc-SNae).  Therefore observations of sufficiently young associations of
massive stars, as the Cygnus region (Kn\"odlseder et al. \cite{Kn02}),
allow us in principle to disentangle WR wind and cc-SNae contributions.

Recently, Smith (\cite{Smith04a}) has reported a first detection of the
1.117 and 1.332 MeV gamma-ray lines attributed to the radioactive decay of
$^{60}$Fe from the inner Galaxy region, using the Reuven Ramaty High Energy
Solar Spectroscopic Imager (RHESSI). After a more deeper analysis, he
revised his first flux estimates downwards, to a marginal 2.6 sigma
detection level (Smith, \cite{Smith04b}). The observed flux in each of the
$^{60}$Fe line amounts to about 10 \% of the \al flux, slightly smaller
than the flux ratio predicted by Timmes et al. (\cite{Ti95}) from
calculations of type II supernovae\footnote{The mass of the progenitors of
  type II supernovae lies between 11\msun and 40\msun, excluding WR stars
  in the case of non--rotating models.}  (hereafter SNII) nucleosynthesis
by Woosley \& Weaver (\cite{WW95}).  Taken at face, such an agreement would
mean that all the galactic \al could originate from the SNII. However, as
discussed in Prantzos (\cite{Pr04}), more recent models of the evolution
and nucleosynthesis of cc-SNae such as those of Rauscher et
al. (\cite{Ra02}) or those of Limongi \& Chieffi (\cite{Li03}), predict
much higher $^{60}$Fe/\al line flux ratios. These models seem thus to leave
more room for sources, such as the WR stellar winds, ejecting \al and not
$^{60}$Fe.  According to Prantzos (\cite{Pr04}), with these new stellar
yields, the RHESSI results can be recovered provided that at least half of
\al originates from WR stellar winds. There are at least three other
arguments supporting some additional contributions to that of the
supernovae.  Firstly, the new supernovae yields appear too low to allow an
injection rate of 2 M$_\odot$ of \al per Myr (Prantzos
\cite{Pr04}). Secondly the observation of a bump of emissivity at 1.809 MeV
in the direction of the Cygnus region, a region with no sign of recent
supernova activity, also points towards the existence of sources such as
the WR stars, that eject \al even before the first supernovae appear
(Kn\"odlseder et al. \cite{Kn02}). Finally, from the point of view of
stellar evolution models, \al enrichment of the interstellar medium by WR
stars is an inherent and unavoidable consequence of the most massive stars
evolution if their observed properties are to be reproduced (Meynet et
al. 1997).  An important contribution of WR stars to the galactic \al is
therefore unavoidable.\\
 
The purpose of the present work is to provide yields from WR stellar winds
based on recent grids of rotating stellar models (Meynet \& Maeder
\cite{paperX}, \cite{paperXI} hereafter paper X and XI respectively).  In
these previous two papers we have investigated the effects of rotation on
the formation of the WR stars at different metallicities. We showed that
the inclusion of rotation makes the formation of WR stars easier, thus
helping to reproduce the observed variations of the number ratio of WR to
O--type stars as a function of the metallicity, of the ratio of WC to WN
stars for metallicities below about $Z$ = 0.020 (solar metallicity), as
well as the observed variation with the metallicity of the number fraction
of type Ibc supernovae with respect to SNII (Prantzos \& Boissier
\cite{Pr03}). Furthermore, models with rotation are able to account for the
small (although significant) fraction of WR stars presenting at their
surface both H-- and He--burning products (see e.g. Crowther et
al. \cite{Cr95}). None of these observed features can
be fitted by non--rotating stellar models computed with recent mass loss
rates prescriptions (see Sect.~2).

Despite the fact that the rotating models well reproduce the mentioned
above observed features as well as the observed surface enrichments
(Heger \& Langer \cite{he00}; Meynet \& Maeder \cite{paperV}, hereafter
paper V) and the number ratio of blue to red supergiants in the Small
Magellanic Cloud (Maeder \& Meynet \cite{MMVII}; hereafter paper VII), the
rotating models have some difficulties in accounting for the observed
fraction of WC to WN stars at high metallicity. Typically, rotating models
seem to underestimate this ratio at twice the solar metallicity by about a
factor of two. In paper XI we attributed this remaining discrepancy as due
to one (or both) of the following two causes:\\ \indent 1) The
incompleteness of the observed sample in high metallicity regions might
bias the result. This seems already to be the case at solar metallicity
where according to Massey \& Johnson (\cite{MJ98}) the observed WC/WN ratio
is overestimated. The less luminous WC stars are more difficult to detect
than the more luminous WN stars, especially in regions where the extinction
may be high. In contrast, in the Magellanic Clouds, where the models well
reproduce the observed ratio, the stars have nearly all the same
metallicity, are at the same distance and the internal extinction is low.\\
\indent 2) The mass loss rates during the post H--burning WN phase are
underestimated. This would increase the WC phase duration at the expense of
the WN phase.  Let us note that even if the mass loss rates would be
underestimated in the post H--burning WN phases, this would not deeply
modify the present results as far as \al is concerned.

Despite the above remaining difficulty, we think that the present models
can sufficiently well account for the observed properties of massive stars
to allow an acceptable estimate of their \al yields.  In Sect.~2 we briefly
recall the physical ingredients of the models. The effects of rotation on
\al production by WR stars at various metallicities are discussed in
Sect.~3. A small subset of the present models for the
metallicities at $Z$ = 0.020 and 0.040 were already briefly discussed in
Vuissoz et al. (\cite{Vu04}). An estimate of the global contribution of WR
stars to the \al present in the Milky Way is given in Sect.~4, while
Sect.~5 presents the main conclusions of this work.


\section{Physical ingredients of the models}

The grid of models presented here was computed with the Geneva stellar
evolution code. The physical ingredients are essentially the same as in
paper X and XI, except for the inclusion of the Ne--Na and Mg--Al reaction
chains and the omission of the effects of the wind anisotropies induced by
rotation (see below). Let us briefly recall the main points relevant for
\al production:
\begin{itemize} 
\item The initial compositions are adapted for the different metallicities
considered here.  For a given metallicity $Z$ (in mass fraction), the
initial helium mass fraction $Y$ is given by the relation $Y= Y_p + \Delta
Y/\Delta Z \cdot Z$, where $Y_p$ is the primordial helium abundance and
$\Delta Y/\Delta Z$ the slope of the helium--to--metal enrichment law. We
use the same values as in paper VII {\it i.e.} $Y_p$ = 0.23 and $\Delta
Y/\Delta Z$ = 2.5.  For the metallicities $Z$ = 0.004, 0.008 and 0.040
considered in this work, we thus have $X$ = 0.757, 0.744, 0.640 and $Y$ =
0.239, 0.248, 0.320 respectively.  For the mixture of the heavy elements,
we adopt the same mixture as the one used to compute the opacity tables for
solar composition (Iglesias \& Rogers \cite{Iglesias}).

\item Since mass loss rates are a key ingredient of the models in the mass
range considered here, let us recall the prescriptions used.  The changes
of the mass loss rates $\dot{M}$ with rotation are taken into account as
explained in Maeder \& Meynet (\cite{MMVI}; hereafter paper VI).  As
reference, we adopt the mass loss rates of Vink et al. (\cite{Vink00};
\cite{Vink01}) who account for the occurrence of bi--stability limits which
change the wind properties and mass loss rates.  For the domain not covered
by these authors we use the results by de Jager et al. (\cite{Ja88}). For
the non--rotating models, since the empirical values for the mass loss
rates are based on stars covering the whole range of rotational velocities,
we must apply a reduction factor to the empirical rates to make them
correspond to the non rotating case. We used a value of 0.85, as in paper
VII.  During the WR phase we use the mass loss rates by Nugis \& Lamers
(\cite{NuLa00}). These rates, which account for the clumping effects in the
winds, are smaller by a factor 2--3 compared to the ones used in our
previous non--rotating ``enhanced mass loss rate'' stellar grids with
predictions of \al yields (Meynet et al. \cite{Me97}).

\item During the non--WR phases of the present models, we assumed that the
mass loss rates depend on the initial metallicity as $\dot
M(Z)=(Z/Z_\odot)^{1/2} \dot M(Z_\odot)$ (Kudritzki \& Puls \cite{KP00};
Vink et al. \cite{Vink01}), while during the WR phase we assumed no
metallicity dependence.  At $Z=0.040$, we also computed a series of models
with metallicity dependent mass loss rates during the WR phase. According
to Crowther et al. (\cite{Cro02}), mass loss rates during the WR phase may
show the same metallicity dependence as the winds of O--type stars, {\it
i.e.} scale with $\sim (Z/Z_\odot)^{1/2}$.

\item A moderate overshooting is included in the present rotating and
non--rotating models. The radii of the convective cores are increased with
respect to their values obtained by the Schwarzschild criterion by a
quantity equal to 0.1 H$_{\rm p}$, where H$_{\rm p}$ is the pressure scale
height estimated at the Schwarzschild boundary.

\item The effect of rotation on the transport of the chemical species and
of the angular momentum are accounted for as in papers VII and VIII (Meynet
\& Maeder \cite{paperVIII}).  All the models were computed up to the end of
the He--burning phase.

\item As initial rotation, we have considered a value equal to 300 \kms on
the ZAMS for all the initial masses and metallicities considered. At solar
metallicity, this initial value produces time averaged equatorial
velocities on the main sequence (hereafter MS) well in the observed range,
{\it i.e.} between 200 and 250 \kms. At low metallicities this initial
rotational velocity corresponds also to mean values between 200 and 250
\kms on the MS, while at twice the solar metallicity, the mean velocity is
lower, between 160 and 230 \kms (see Table 1 in paper XI). Presently we do
not know the distributions of the rotational velocities at these non--solar
metallicities and thus we do not know if the adopted initial velocity
corresponds to the average observed values.  It may be that at lower
metallicities, the initial velocity distribution contains a larger number
of high initial velocities (Maeder et al.~\cite{MG99}), in which case the
effects of rotation described below would be underestimated at low
metallicity.

\item The Ne--Na and Mg--Al nuclear reaction chains have been included in
addition to the usual nuclear reactions for the H-- and He--burning phases.
The long--lived ($t_{1/2} = 7.05 \times 10^5$ yr) $^{26}{\rm Al}^g$ ground
state is considered as a separate species from its short--lived ($t_{1/2} =
6.35$ s ) $^{26}{\rm Al}^m$ isomeric state. In the following, for purpose
of simplicity, the long--lived $^{26}{\rm Al}^g$ ground state will be
denoted by the symbol \al. The nuclear reaction rates are taken from the
NACRE compilation (Angulo et al. \cite{Ang99}). Arnould et
al. (\cite{Arn99}) have studied in a simple parametric model the
uncertainties on the abundances of some elements due to the uncertainties
of the nuclear reaction rates. For what concerns \al they reach the
interesting conclusion that the large uncertainties on the $^{26}{\rm
Al}(p,\gamma)^{27}{\rm Si}$ reaction rate scarcely affect the abundance of
\al, considering that ``{\it even the highest NACRE proton capture rates
are not fast enough for leading to a substantial destruction of the two Al
isotopes by the time H is consumed}'' (Arnould et al. \cite{Arn99}) .
\end{itemize}
Let us finally recall that the wind anisotropies induced by rotation were
neglected. This last choice appears justified in view of the results
obtained in paper X. Indeed for the initial velocities considered
($\upsilon_{\rm ini} = 300$ \kms), the effects of the wind anisotropies
have been shown to be very small. Let us however emphasize that this is not
true for higher initial velocities (Maeder \cite{Ma02}).

\section{Effects of rotation on \al production by WR
  stars}
\begin{figure}[h]
\resizebox{\hsize}{!}{\includegraphics[width=8cm,height=8cm,angle=-90]{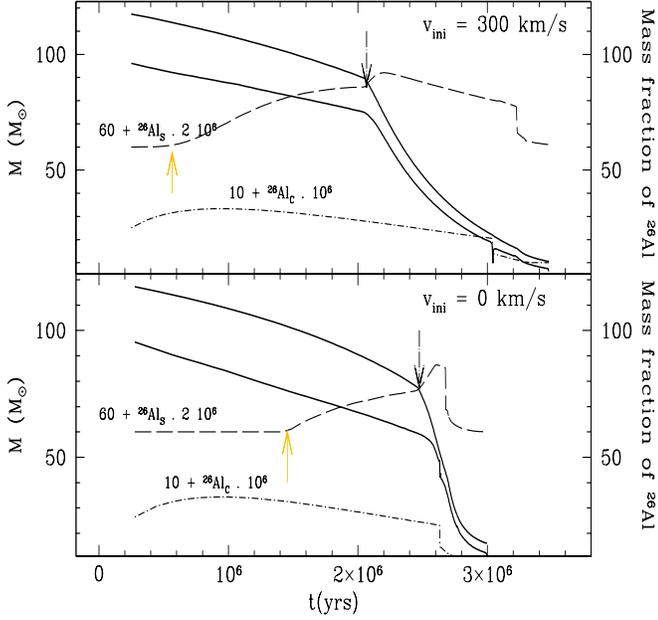}}
\caption{Evolution as a function of time of surface ($^{26}{\rm Al}_s$) and
  central ($^{26}{\rm Al}_c$) mass fraction of \al for both rotating (upper
  panel) and non--rotating (lower panel) 120 ${\rm M}_{\odot}$ stellar
  models at $Z$ = 0.020. Light dashed lines show the scaled values of the
  mass fractions 60 + $^{26}{\rm Al}_s \times 2~10^6$ and 10 + $^{26}{\rm
  Al}_c \times 10^6$. Bold solid lines mimic a Kippenhahn like diagram, and
  show the evolution of the mass of the convective core and that of the
  total stellar mass as a function of time. The down--going arrows indicate
  the beginning of the strong wind phase, and the up--going arrows indicate
  the beginning of the surface \al enrichment.}
\label{figure1}
\end{figure} 

\begin{figure}[h]
\resizebox{\hsize}{!}{\includegraphics[width=8cm,height=8cm]{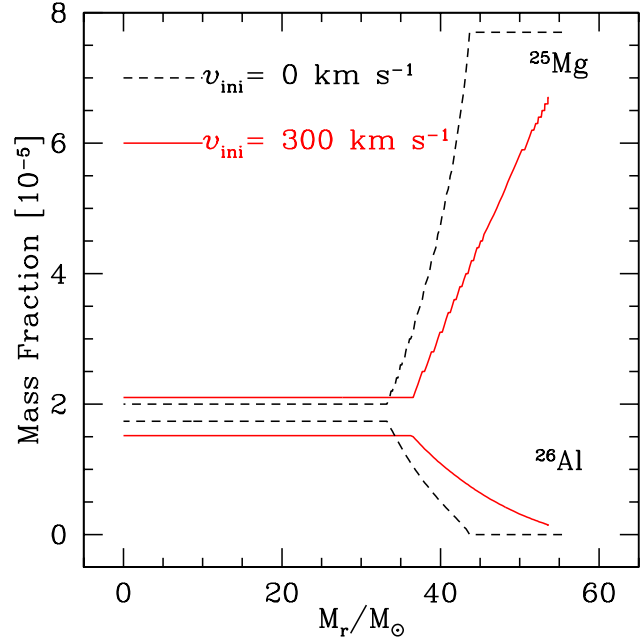}}
\caption{Variation of the mass fractions of $^{25}$Mg and \al as a function
of the lagrangian mass inside a rotating and non--rotating 60 M$_\odot$
stellar model at $Z$ =0.020 during the core H--burning phase. In both
models the mass fraction of hydrogen in the convective core is equal to
0.4.}
\label{figure2}
\end{figure} 

\al is mainly produced by proton capture on $^{25}{\rm Mg}$ seeds in the
  convective core during the central Hydrogen Burning phase (see
  Fig.~\ref{figure1} and Fig.~\ref{figure2}). In non--rotating stars, part
  of the \al produced in the core, depending on the initial stellar mass,
  may appear at the surface due to the removal of the envelope by stellar
  winds. From this point on, stellar winds enrich the interstellar medium
  in \al.  Including rotation somewhat modifies this picture, as we will
  show in the following sections.\\ Before entering into the details of the
  effects of rotation on \al release by WR stars, let us briefly recall
  those effects of rotation on massive star evolution which may be relevant
  in the context of \al nucleosynthesis :
 
\begin{enumerate}
\item Rotation allows diffusion of the chemical species from the convective
  core to the radiative envelope and vice versa; this results in changes of
  the surface abundances already during the Main Sequence phase. The
  surface enrichments given by the models are generally in good agreement
  with the observed ones (see Heger \& Langer \cite{he00}; Meynet \& Maeder
  \cite{paperV}; Maeder \& Meynet \cite{MMVII})
\item Rotation increases mass loss (see Sect~2). More precisely a rotating
star of given initial mass and position in the HR diagram undergoes a
higher mass loss than a non--rotating star (paper VI).
\item Diffusion of elements actively contributing to the nuclear energy
  production inside the star together with enhanced mass loss allow massive
  stars to enter the WR phase at an earlier stage of their evolution. As a
  consequence, the WR phase lasts for longer than in the case of
  non-rotating stars.
\item Finally, rotation decreases the minimum initial mass for a single
star to enter the WR phase (see Table~2).
\end{enumerate} 

These effects, when accounted for in stellar models, much improve the
agreement between theory and observations for what concerns the
characteristics of the WR populations at different metallicities (see paper
X and XI).  Through these different processes, rotation also modifies the
quantity of \al ejected by massive stars undergoing strong winds.

In Fig.~\ref{figure1} to Fig.~\ref{figure3}, we present the evolution of
the \al abundance in test models of 60 and 120 \msun at different
metallicities. In Fig.~\ref{figure4} and Table 1, we provide the wind
ejected mass of \al
$$
M_{26}^W[M_i,Z_i] = \int^{\tau(M_i,Z_i)}_0  {^{26}{\rm Al}_S} (M_i,Z_i,t)
 |{\dot M} (M_i,Z,t) | dt \eqno (1)
$$ 
for models with $M_i \in$ [25 \msun;120 \msun] and metallicities from Z =
0.004 to Z = 0.04.  Here we do not take the radioactive decay of \al into
account, provided that it is the production rate of \al that is needed to
estimate the contribution of WR stars to the present galactic mass of \al.

The effects of rotation on \al nucleosynthesis are the followings:

\begin{enumerate}
\item 
  As can be seen from Fig.~\ref{figure1}, rotation slows down the decrease
  in mass of the convective core during central hydrogen burning
  phase. This results from the fact that rotational mixing continuously
  brings fresh H--fuel into the core. The regions where \al is synthesized
  are closer to the surface, making it easier for the mixing to connect
  both regions, thus favouring an increase of the quantity of \al ejected.
  One notes also from Fig.~\ref{figure1} that whether rotation is taken
  into account or not, the maximum content of \al inside the convective
  core is about the same in both cases, and it is reached at 1
  Myr. Afterwards, the amount of \al slowly decreases as its production via
  the reaction $^{25}{\rm Mg}(p,\gamma)^{26}{\rm Al}$ is not sufficient
  anymore to compensate its reduction due to radioactive decay.\\
\begin{table*}[t]
\begin{center}
\caption{Wind ejected mass of \al in units of $10^{-4} {\rm M}_{\odot}$ for
  both rotating and non--rotating models at the different metallicities
  considered. Bold values are for models computed with metallicity
  dependent mass loss during the WR phase. }
 \begin{tabular}{c|c|cccccc}
 \hline
 \hline
\footnotesize{Z} & \footnotesize{$\upsilon_{ini}~~(km/s)$} &\footnotesize{${\rm M}_i$ = 120 \msun} & \footnotesize{${\rm M}_i$ = 85 \msun} & \footnotesize{${\rm M}_i$ = 60 \msun} & \footnotesize{${\rm M}_i$ = 40 \msun} & \footnotesize{${\rm M}_i$ = 30 \msun}  & \footnotesize{${\rm M}_i$ = 25 \msun}\\ 
 \hline
& \footnotesize{300} & \footnotesize{25.74} & \footnotesize{13.07} & \footnotesize{7.567}  & \footnotesize{1.980}  & \footnotesize{} & \footnotesize{0.962}\\
\footnotesize{0.04} & \footnotesize{300} &  \footnotesize{\bf 22.44} &  \footnotesize{\bf
  11.87} &\footnotesize{\bf 7.2056} & \footnotesize{\bf 2.029}
&\footnotesize{} &  \footnotesize{}  \\   
 & \footnotesize{0} & \footnotesize{21.62} & \footnotesize{} & \footnotesize{3.021}  & \footnotesize{}  & \footnotesize{} & \footnotesize{0.029}\\\hline
\footnotesize{0.02} & \footnotesize{300} & \footnotesize{12.29} & \footnotesize{5.83} & \footnotesize{2.181}  & \footnotesize{}  & \footnotesize{} & \footnotesize{0.345}\\
& \footnotesize{0} & \footnotesize{5.743} & \footnotesize{} &
\footnotesize{1.304}  & \footnotesize{}  & \footnotesize{} &
\footnotesize{0.001}\\\hline
\footnotesize{0.008} & \footnotesize{300} & \footnotesize{2.566} & \footnotesize{} & \footnotesize{0.297}  & \footnotesize{0.210}  & \footnotesize{0.096} & \footnotesize{}\\\hline
\footnotesize{0.004}& \footnotesize{300} & \footnotesize{0.701} & \footnotesize{} &
\footnotesize{0.129}  & \footnotesize{0.029}  & \footnotesize{0.002} &
\footnotesize{}\\\hline

\end{tabular}
\end{center}
\end{table*}

\begin{figure*}
\resizebox{\hsize}{!}{\includegraphics[angle=-90]{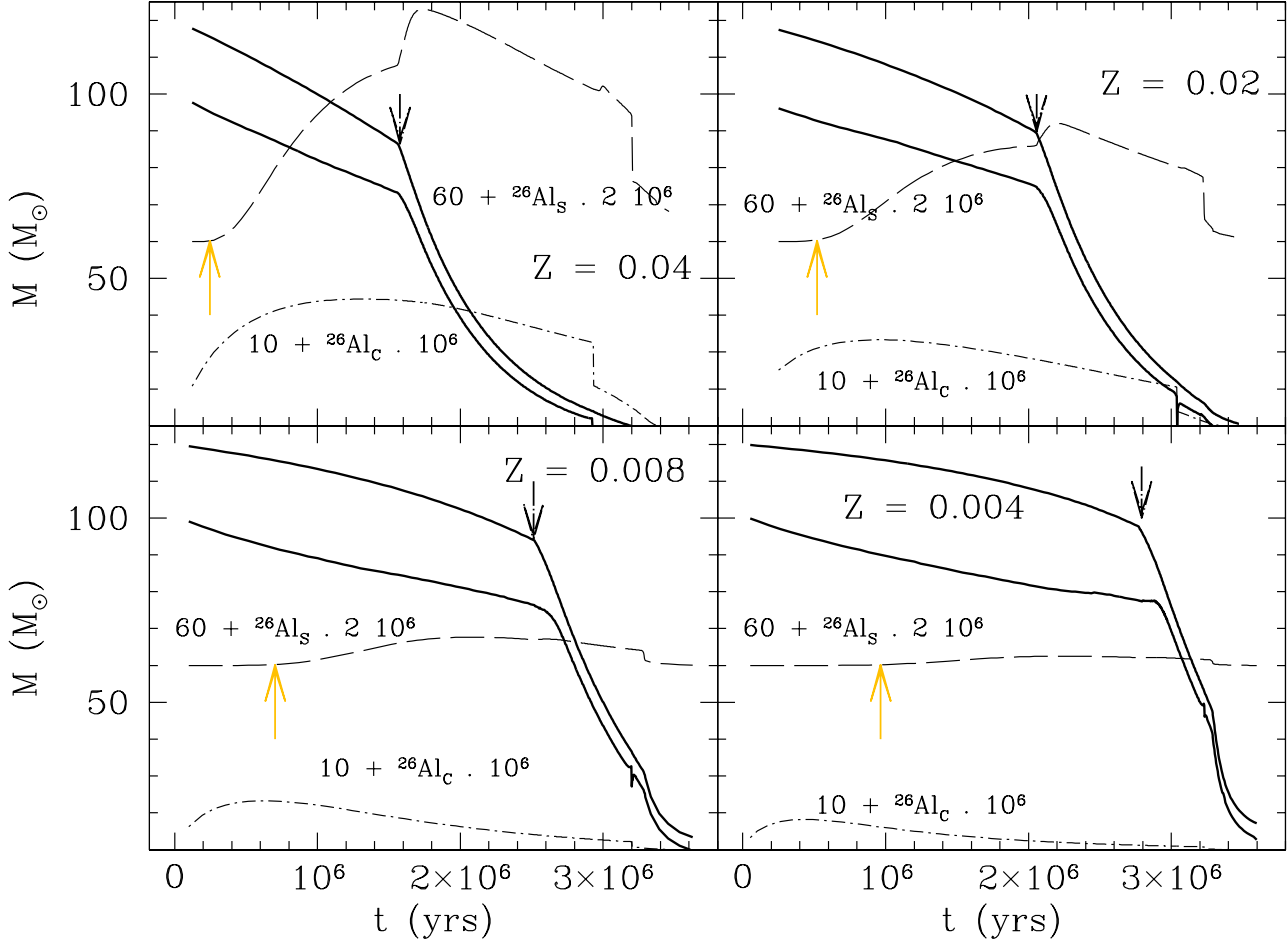}}
\caption{Same as Fig. 1 for rotating models of a 120 ${\rm M}_{\odot}$ star
at four different metallicities, as indicated on the plots.}
\label{figure3}
\end{figure*}

\item 
  Due to rotational mixing, the \al surface abundance begins to change at a
  much earlier time in the rotating model. This can also be seen from
  Fig.~\ref{figure2} which shows the variations inside rotating and
  non--rotating 60 M$_\odot$ stellar models of the mass fractions of
  $^{25}$Mg and \al at the middle of the core H--burning phase.  While the
  abundances of these two elements are very similar, the main differences
  arise in the radiative envelope: in the non--rotating model, above the
  transition zone left over by the recession of the convective core, the
  abundances present a flat profile at a level corresponding to the initial
  abundance (which is of course 0 for \al). Instead in the rotating model,
  the inwards diffusion of $^{25}$Mg and the outwards diffusion of \al
  smooth the profiles and produce changes of the surface abundances while
  mass loss has not yet uncovered layers previously pertaining to the
  convective core.\\

\item
  From Fig.~\ref{figure1} and Fig.~\ref{figure3}, it appears that most of
  the \al is ejected while the star is at the beginning of the WR
  phase. Indeed at this stage, the layers removed by the stellar winds are
  heavily loaded in \al and the mass loss rate is high. When layers
  processed by He--burning appear at the surface, the quantity of \al much
  decreases, because \al is destroyed (mainly by neutron capture) at the
  beginning of the core He--burning phase.  Since rotating stars enter the
  WR phase at an earlier evolutionary stage than non--rotating ones (see
  Fig.~\ref{figure1}), the period during which \al is released into the ISM
  is longer, and the net yield of this nuclide is also increased when
  rotation is taken into account.  This appears clearly in Table 1, where
  we show the total wind ejected mass of \al for the computed models.\\

\item
  The lowering of the minimum mass for a star to enter the WR phase
  contributes to enhance considerably the total contribution of WR stars to
  the global galactic content of \al. Less massive stars, favoured by the
  IMF, will now contribute to the \al enrichment of the medium through
  their winds, whereas their non--rotating counterparts could only
  contribute through the supernova explosion (see Sect. 4).
\end{enumerate} 

\subsection{Relative effects of mass loss and rotation on \al}

 Values listed in Table 1 show that for the higher metallicity models the
 \al mass ejected is increased by a factor of about 2.5 for typical WR
 progenitors with ${\rm M}_{\rm init} \sim$ 40--85 \msun. For the 25 \msun
 model, the \al mass released by rotating models is 33 times higher than
 the one obtained for non--rotating ones, mainly because in the latter
 case, these stars do not enter the WR phase and contribute very few
 through their winds to the \al enrichment of the interstellar medium.  In
 the lower mass stars, the wind during the WR phase is weaker than for the
 more massive ones, and rotation--induced mixing is the main mechanism
 driving the evolution of the surface abundance patterns. On the contrary,
 in the mass range ]60 \msun; 120 \msun], the surface abundances during the
 WR phase essentially result from the efficiency of the wind to peel off
 the star, so that the effects of rotation--induced mixing do not appear
 that strikingly when comparing rotating and non--rotating models (see also
 Fig.~\ref{figure4}).

It is interesting to note from Fig.~\ref{figure4} that the yields from
Meynet et al. (\cite{Me97}), computed for non-rotating models with enhanced
mass loss are lower than the ones obtained for rotating models with lower
mass loss rates compatible with the most recent observational determination
(see Sect.~2). This again points out the importance of rotation, non solely
in increasing mass loss, but mainly in allowing for transport of the
chemicals throughout the stellar interiors.

\medskip
\subsection{Metallicity dependence of \al yields}

 As shown in previous papers (Meynet et al. \cite{Me97}, Vuissoz et
  al. \cite{Vu04}), the amount of \al both synthesized and expelled by a WR
  star increases with metallicity. This is already the case for
  non-rotating models, and rotation adds to this effect as can be seen in
  Figs. 3 and 4.

\begin{figure}
\resizebox{\hsize}{!}{\includegraphics{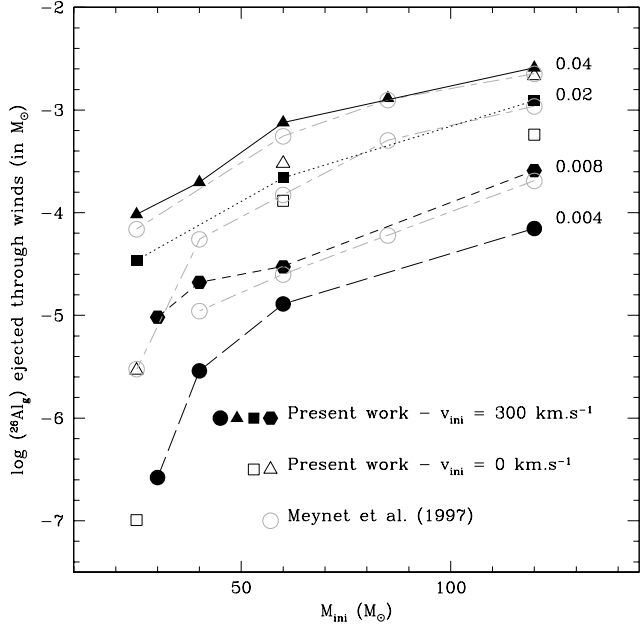}}
\caption{\al yields, as listed in Table~1, as a function of initial stellar
  mass and metallicity. Filled symbols are for rotating models with
  $\upsilon_{ini} = 300 ~{\rm km}.{\rm s}^{-1}$, open black squares and
  triangles are for non--rotating models and open grey circles connected by
  long--short dashed lines are from Meynet et al. (1997). The corresponding
  metallicity is indicated on the right part of the figure.}
\label{figure4}
\end{figure}

  In Fig.~\ref{figure3}, we present the evolution of the surface and center
  mass fraction of \al as a function of time for rotating 120 M$_\odot$
  stellar models at various metallicities. One sees that the higher the
  metallicity, the larger the amount of \al produced in the convective core
  and the earlier \al appears at the surface.  Thus for a given initial
  mass and rotation, the mass of \al ejected is greater at higher
  metallicity. This is also well illustrated in Fig.~\ref{figure4}. From
  the present models one expects a relation of the type $Y(^{26}{\rm Al},
  M_{\rm ini}) \propto Z_{\rm ini}^{\beta}$ between the yield and the
  metallicity.  Here, the value of $\beta$ depends on both initial mass and
  rotation. Typically from the rotating stellar models with M$_{\rm ini}$ =
  40, 60 and 120 \msun, one obtains $\beta=1.4$, 1.8 and 1.4 respectively
  in the metallicity range from $Z$=0.008 to 0.040.  For the non--rotating
  60 and 120 M$_\odot$ and in the metallicity range from 0.02 to 0.04, one
  obtains values of $\beta$ equal to 1.2 and 1.9 respectively.  As can be
  seen from these numerical examples, the concomitant effects of rotation
  and mass loss are intricate and it is difficult to draw more general
  trends for what concerns the sensitivity of the yields on the initial
  metallicity.

\subsection{Effects of metallicity dependent mass loss during the WR phase}

  \indent For twice the solar metallicity, we computed models using
  Crowther et al. (\cite{Cro02}) prescription for the mass loss rates
  during the WR phase. For this metallicity, the derived mass loss rates
  are multiplied by a factor $\sim 1.4$. The resulting wind ejected masses
  of \al are listed in bold face in Table 1.  For the most massive stars,
  the masses of ejected \al do not differ by more than 13 \% when the
  metallicity dependent mass loss rates are used. At first sight this may
  appear surprising.  However the most massive stars enter the WR phase
  while central H--burning is still going on.  In that case a stronger mass
  loss produces a more important reduction of the convective H--burning
  core, accompanied by a severe reduction of the central temperature. The
  end of the H--burning phase is then lengthened, and thus the \al present
  inside the convective core has more time to decay {\it in situ} before
  being expelled.\\ The changes are however small and therefore one can
  conclude that the use of metallicity dependent mass loss rate does not
  affect significantly the results.\\

\subsection{Yields dependence on initial velocity}
Finally let us recall that in Table~1, the yields in \al for rotating
models are given for an initial rotation velocity on the ZAMS of 300
\kms. This value corresponds to an average velocity during the MS phase
between 200 and 250 \kms, {\it i.e.} well in the observed velocity range
for the OB Main Sequence stars at solar metallicity. In order to estimate
the effects of a higher initial velocity, we have computed, for solar
metallicity, a 60 M$_\odot$ stellar model with $\upsilon_{\rm ini}$ = 500
km s$^{-1}$. The quantity of \al ejected by the stellar winds is in this
case of 2.6 10$^{-4}$ M$_\odot$, which is less than 20\% higher than the
yields obtained from the model with $\upsilon_{\rm ini}$ = 300 km
s$^{-1}$. Thus at least in this particular case, the yields do not appear
to be very sensitive to a change of the initial velocity in the range
between 300 and 500 km s$^{-1}$.

\section{Contribution of WR stars to the \al galactic
  content}

Let us first consider a simple way to estimate the global contribution of
WR stars to the present day \al content of the Galaxy. According to van der
Hucht (\cite{VdH02}), the galactic distribution of WR stars shows a
projected surface density of 2.7 WR stars per kpc$^2$ in the solar
neighbourhood. Extrapolating this estimate of the surface density to the
whole Galaxy gives a total number of WR stars in the Galaxy of the order of
about 2000. Since the average lifetime of WR stars is $\sim 1$ Myr
(e.g. paper XI) this gives a galactic frequency $f_{\rm WR} \sim 2000$
Myr$^{-1}$. Let us note that these numbers are likely to be lower limits,
since the star formation rate and the number of WR to O-type stars is much
higher in inner galactic regions.

The galactic frequency of WR stars may also be estimated by using a stellar
Initial Mass Function (IMF) and the current death rate of massive stars in
the Milky Way.  Observations of external spiral galaxies indicate that the
expected supernova frequency in our Galaxy is $f_{\rm SN} =
2.5^{+0.8}_{-0.5}$ per century, with 85\% of them coming from massive stars
(Tammann et al.  \cite{Ta94}). We adopt then $f_{\rm SN}=2$ per century for
the death rate of massive stars in the Milky Way. We use a Salpeter
(\cite{S55}) stellar IMF\footnote{In the following, we will also discuss
the effect of using a larger slope ($x=1.7$) that could be more appropriate
for massive stars when accounting for binarity corrections (Scalo 1986;
Kroupa et al. 1993).}, $\Phi(M) \propto$ M$^{-(1+x)}$ with $x=1.35$.  The
WR frequency is then given by:
$$
f_{\rm WR}\> =\> f_{\rm SN}\,
{{\int^{M_{\rm UP}}_{M_{\rm LWR}} \Phi(M) dM}\over{\int^{M_{\rm UP}}_
{M_{\rm LSN}} \Phi(M)dM}}\>\sim\>
f_{\rm SN}\, \Biggl({M_{\rm LWR}\over M_{\rm LSN}}\Biggr)^{-x}, \eqno (2)
$$ 
where $M_{\rm LSN}$ and $M_{\rm LWR}$ are the lowest masses of stars
that may become supernovae and WR, respectively. The former is $M_{\rm
LSN}\sim$ 8 M$_\odot$, while the latter depends on metallicity and adopted
mass loss rates. According to our calculations with rotation $M_{\rm
LWR}\sim$ 21 M$_\odot$ for $Z = 2 Z_\odot$ (see Table 2). We consider here
the high metallicity models since, most of the WR stars are found in the
inner regions of our Galaxy where the metallicity is higher (see below).
Taking these values into account one obtains $f_{\rm WR}\sim 0.2 ~f_{\rm
SN}$, i. e. a WR frequency of 0.4 per century or 4000 WR stars per Myr.  On
the other hand, from the yields of Table 1 and our adopted IMF, an average
WR yield of $\langle M_{26}^W\rangle \sim 3.4\, 10^{-4}M_\odot$ can be
defined for $Z= 2 Z_\odot$ stars.

The collective production rate of \al in the Galaxy by WR stars can
then be expressed as 
$$
{\dot M}_{26} \sim 1.4 {{\langle M_{26}^W\rangle}\over{3.4\,10^{-4}}}
{{f_{\rm WR}}\over{4000}}~~~~{\rm M}_{\odot} \cdot {\rm Myr}^{-1}. \eqno (3)
$$
\\ In the case of a stationary regime as it assumed to be the case in the
Milky Way, the galactic mass of \al does not vary with time, so that at a
given time $t$, the production ($P_{26}(t)$) and destruction ($D_{26}(t)$)
rates of \al are equal.  The destruction rate is easy to evaluate,
considering the evolution of the \al mass solely due to its decay , e.g. $$
{\rm M}_{26}(t) = {\rm M}_{26}(t_0) \cdot e ^{-\frac{t}{\tau}} \Rightarrow
D_{26} = \frac{{\rm M}_{26}}{\tau}.$$ As $P_{26}(t) = D_{26}(t)$, the
production rate of \al is then given by :
$$P^{gal}_{26} = D^{gal}_{26} = {\dot M}^{gal}_{26} = \frac{{\rm
    M}^{gal}_{26}}{\tau}~~{\rm M}_{\odot}~{\rm per}~{\rm unit}~{\rm
    time}. \eqno (4)$$ The total mass of \al originating from WR stars
    present in the Galaxy is related to the production rate of \al by these
    stars through Eq.~(4). The mean lifetime of \al being of the order of
    1.04 Myr, we finally have:
$${\rm M}^{gal}_{26} = \tau_{26} \times {\dot M}^{gal}_{26} \approx {\dot
  M}^{gal}_{26}~~ {\rm M}_{\odot}. \eqno (5)$$ Thus Eq.~(3) (and Eq.~(6) in
  the following) in addition to give the injection rate of \al by WR stars,
  can also be used to evaluate the total mass of \al originating from WR
  stars. According to our rough estimate from Eq. (3), the mass of \al
  contributed by WR stars amounts to 45 \% and even 70 \% of the
  observationally inferred mass of about 1.8 \msun (Kn\"odlseder
  \cite{Kn97}).\\

In a more formal way, the \al production rate (in solar mass per unit time)
from galactic WR stars may be evaluated as 
$$
{\dot M}_{26} = \int^{\rm R}_0 2\pi r \sigma(r)
\int^{120}_{M_{\rm LWR}(r)} \Phi(M) M_{26}^W[M_i,Z_i(r)] dM~ dr, \eqno (6)
$$ 
where $\sigma(r)$ is the Star Formation Rate (hereafter SFR) per unit
surface in the Galaxy given as a function of galactocentric distance $r$
and $M_{26}^W$ is the mass of \al ejected by a star of initial mass $M_i$
and initial metallicity $Z_i(r)$, according to Eq.~(1).\\ To account for
the metallicity dependence of the \al yields and of the lowest WR mass
$M_{LWR}$ in Eq. (6), we use the Z-dependent values given in Table 1 and in
Table~2 respectively.

\begin{table}
\begin{center}
\caption{Minimum mass $M_{LWR}$ for a star to enter the WR phase as a
function of $Z$ and $\upsilon_{ini}$ ({\it from Meynet \& Maeder (2004)})}.
\begin{tabular}{c|c|cccc}
\hline
\hline
 & $\upsilon_{ini}$ & Z & Z  & Z  & Z \\
 & (\kms) & 0.04 & 0.02 & 0.008 & 0.004\\\hline
$M_{LWR}$ & 0 & 29 & 37 & 42 & 52 \\
(\msun) & 300 & 21 & 22 & 25 & 32\\\hline
\end{tabular}
\end{center}
\end{table}
The integration in Eq. (6) requires the following ingredients:

\begin{itemize}

\item{$\bullet$} Normalization of both the SFR $\sigma(r)$ and the IMF
  $\Phi(M)$. We can either rely, as in our first simple estimate, on an
  approximate frequency of SNae events in the Galaxy; in this case, the
  integral $\int^{\rm R}_0 2\pi r \sigma(r)dr \int^{120}_{8}\Phi(M)dM$ is
  normalized to $2\, 10^4$ Myr$^{-1}$ (2 cc-SNae per century in the
  Galaxy).\\ We can also follow Kn\"odlseder (\cite{Kn99}), and normalize
  the SFR so that the total predicted Lyman continuum photon production of
  the Galactic population defined by the IMF matches the observed Lyman
  continuum luminosity Q. The actual value of Q as deduced by McKee \&
  Williams (\cite{McKW97}) is equal to 2.56 $\times 10^{53}$ photons. ${\rm
  s}^{-1}$. In that case, according to Eqs. (18) and (20) of Kn\"odlseder
  (\cite{Kn99}), we will have $\int^{120}_{1} \Phi(M) dM = 1$ and the star
  formation rate $\int^{\rm R}_0 2\pi r \sigma(r) dr$ normalized to
  $\frac{Q}{\int^{120}_8 s_{H}(M) \Phi(M) \tau_{total}(M) dM} $, $s_{H}(M)$
  being the time--averaged Lyman continuum luminosity and $\tau_{total}(M)$
  the total lifetime of a star of initial mass $M$. To evaluate these two
  quantities, we make use of Eqs.~(21) of Kn\"odlseder (\cite{Kn99}), which
  is a rough approximation when taking rotation into account, considering
  that these estimates are for non--rotating models with different mass
  loss prescriptions than the ones used in the present paper. Keeping this
  in mind, we nevertheless found it interesting to estimate the effect of
  normalization on the \al injection rates predicted by Eq. (6), and used
  both SNae and Lyman continuum for this purpose.

\item{$\bullet$} The metallicity gradient plays an important role for
determining the global contribution of WR stars to the galactic \al
enrichment. In order to illustrate quantitatively this point, we performed
three estimates with three different galactic metallicity distributions in
the Galaxy.\\ Our first choice (model $Z1$) is based on the
metallicity gradient deduced from observations of HII regions by Shaver et
al. (\cite{S83}) and on the galactic centre metal abundance obtained by
Najarro (\cite{Na99}).  It corresponds to a metallicity gradient of $d\log
Z / dr = -0.07$ dex.kpc$^{-1}$ (with $Z$(8.5 kpc)=0.02) in the whole region
lying between 1.7 and 15 kpc.  In the inner 1.7 kpc, the metallicity was
taken equal to 0.06 ({\it i.e.}  three times solar). Since we did not
compute models at such a high metallicity, we estimated the \al yields for
metallicities above 0.04 extrapolating those obtained at solar and twice
the solar metallicity.\\ Our second choice (model $Z2$) is based on
the metallicity gradient proposed by Andrievsky et al. (\cite{A04}), which
relies on cepheids observation. They divide the Galaxy into three zones:
Zone I encompasses the region between 4.0 and 6.6 kpc. In this zone
metallicity varies according to the law $\log Z/Z_\odot=-0.128 R_G+0.930$,
$R_G$ being the galactocentric distance expressed in kpc. Zone II extends
from 6.6 to 10.6 kpc and has a metallicity law $\log Z/Z_\odot=-0.044
R_G+0.363$ (for $R_G=8.5$ kpc, $Z\sim Z_\odot$).  The third zone goes from
10.6 to 14.6 kpc and has a metallicity law of the form\footnote{There is an
error in the expression given by Andrievsky et al. (\cite{A04}) for this
zone, that can be corrected considering $\log (Z+0.024)/Z_\odot$ instead of
$\log Z/Z_\odot$.}  $\log (Z+0.024)/Z_\odot=+0.004 R_G+0.256$. For
galactocentric distances inferior to 4 kpc, we take a constant value of the
metallicity equal to 2.62 $\times$ $Z_\odot$.\\ Finally in  model
$Z3$, we consider a much shallower gradient of $d\log Z / dr = -0.04$
dex.kpc$^{-1}$ in the whole region between 4.4 and 12.9 kpc as suggested by
Daflon \& Cunha (\cite{DC03}). For galactocentric distances below 4.4, we
suppose a constant metallicity equal to 0.03 and for galactocentric
distance above 12.9 kpc, we take a constant metallicity equal to 0.013.

These three hypotheses not only account for the various values of the
metallicity gradients found in the literature, but also for the different
views on the metallicity of the galactic centre.  For instance Ram\'irez et
al. (\cite{R00}) and Najarro et al. (\cite{Na03}) find that the mean [Fe/H]
of the Galactic Centre stars is similar to that of the solar neighbourhood
stars, while Najarro (\cite{Na99}) obtains values between 2 and 3 times the
solar metallicity.

\item{$\bullet$} Different estimates can be found in the literature for the
radial distribution of the present Star Formation Rate in the Galaxy
$\sigma(r)$.  We chose to consider two different prescriptions : Paladini
et al. (\cite{P04}), give the distribution of HII regions in our Galaxy as
a function of the galactocentric distance. HII regions are considered to be
good tracers of the SFR, and we suppose that their surface density
distribution is directly proportional to it. Wang \& Silk (\cite{WS94}), on
an other hand, give the SFR per unit surface, based on observations of
various tracers (Lyman continuum photons from HII regions; pulsars;
supernova remnants) and normalized to the solar neighbourhood.

\end{itemize} 
 
Applying Eq. (6) with a Salpeter IMF slope, a SFR from Wang \& Silk
(\cite{WS94}) and a normalization on the rate of SNae events, leads to an
injection rate of ${\dot {\rm M}} \sim 0.3-0.6$ \msun/Myr of \al for the
non--rotating models and to values between 0.9 and 1.3 \msun/Myr for the
rotating models, not far from the order of magnitude estimates performed
above (see Eq. 3).  Rotation, all other physical ingredients being the
same, thus roughly doubles the contribution of the WR stellar winds to the
galactic \al enrichment, as can be seen from values in Table~3. This remark
holds whatever the combination of the physical ingredients entering Eq. (6)
used. In that respect, rotation (for the adopted value for the initial
velocity, $\upsilon_{\rm ini}$ = 300 \kms) has a similar impact on the
global contribution of the WR stars as an enhancement of mass loss rates
(see Meynet et al. \cite{Me97}).
\medskip

\begin{figure}
\resizebox{\hsize}{14cm}{\includegraphics{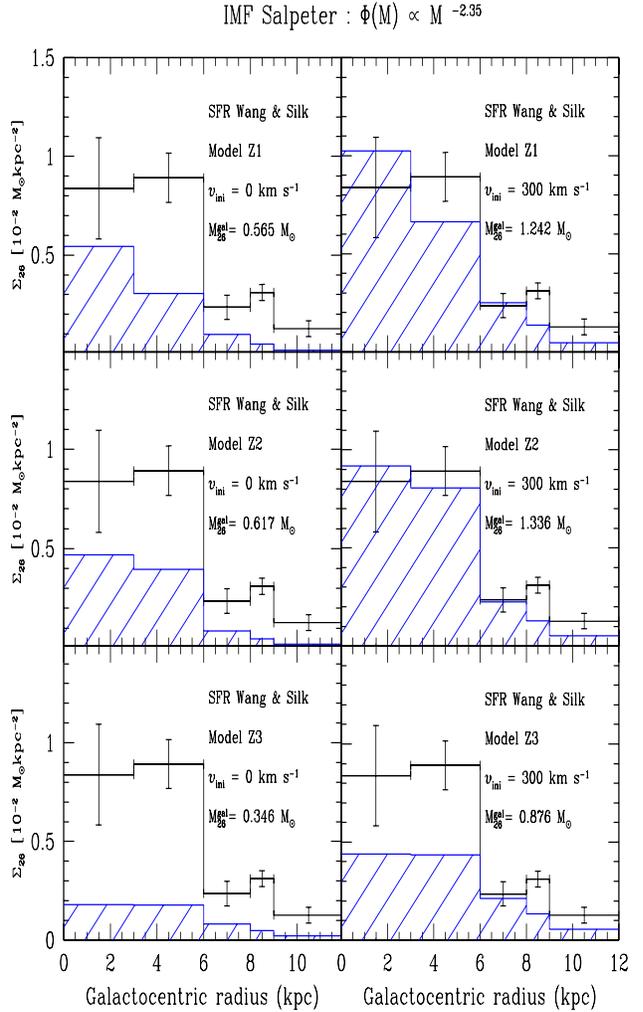}}
\caption{Mass surface density profile of \al (in $10^{-2} \cdot$ \msun
  $\cdot {\rm kpc}^{-1}$ ) as a function of the galactocentric radius for
  an IMF index of 1.35 and a SFR from Wang \& Silk (1994), normalized in
  order to reproduce a rate of 2 SNae events per century in the Galaxy. The
  black histogram with error bars is the profile deduced from COMPTEL data
  analysis (Kn\"odlseder 1997). The dashed histograms are the predicted
  distribution for the prescriptions adopted when using non--rotating (left
  panels) or rotating models (right panels). }
\label{figure5}
\end{figure}

\begin{figure}
\begin{center}
\resizebox{\hsize}{14cm}{\includegraphics{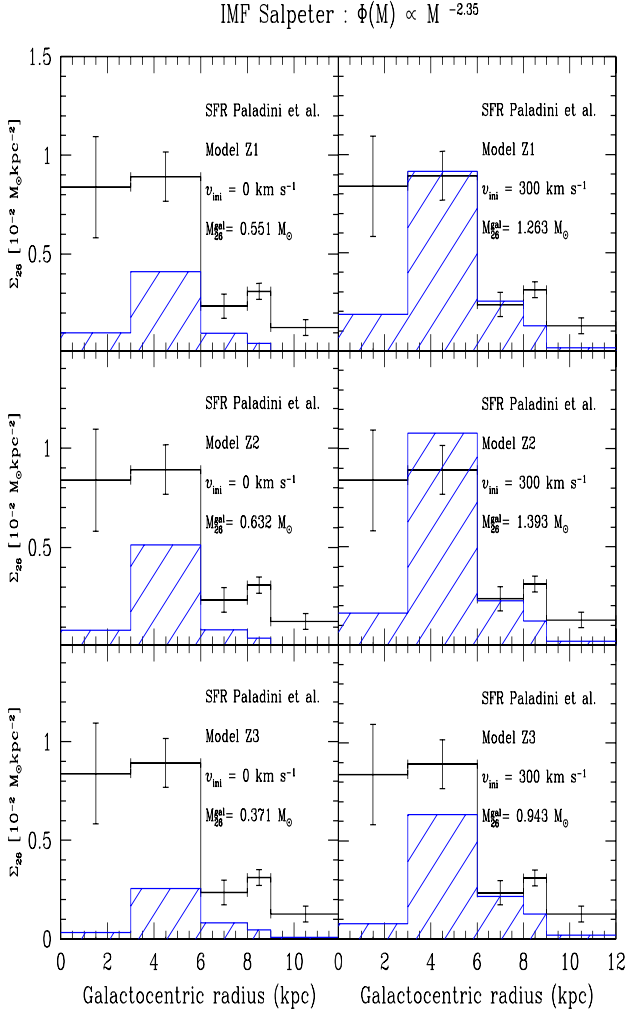}}
\end{center}
\caption{Same as Fig.~5, for a SFR based on Paladini et al. (2004)
  number counts of HII clouds, and $\int^{\rm R}_0 2\pi r \sigma(r)dr
\int^{120}_{8}\Phi(M)dM$ normalized to $2~10^4$ SNae/Myr.}
\label{figure6}
\end{figure}

\begin{figure}
\begin{center}
\resizebox{\hsize}{14cm}{\includegraphics{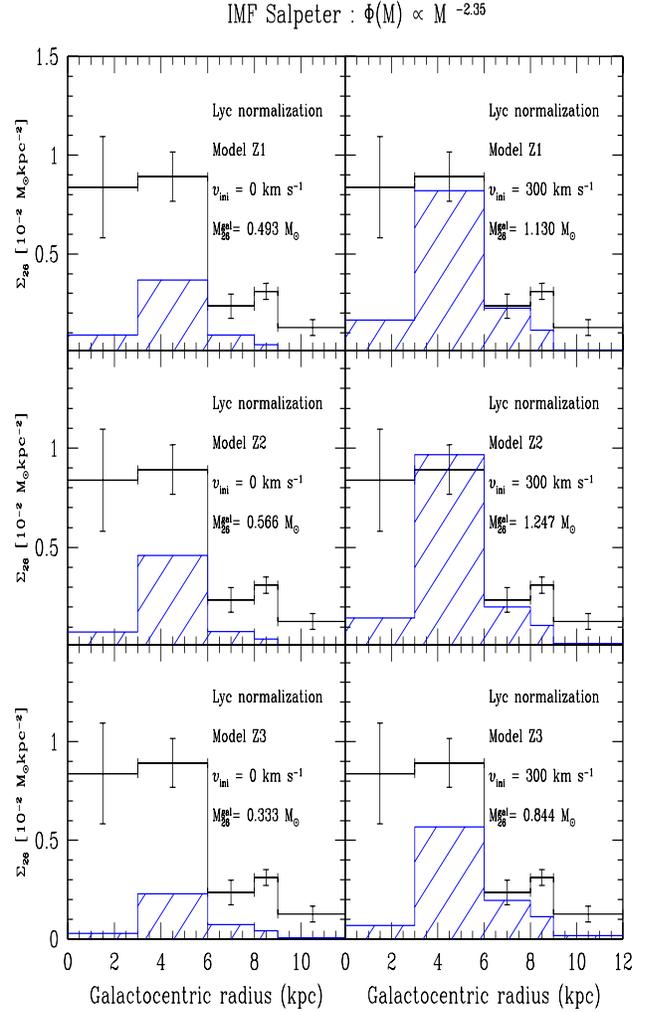}}
\end{center}
\caption{Same as Fig.~5, for a SFR based on Paladini et al. (2004)
  number counts of HII clouds, normalized to the observed Lyman continuum
  luminosity of Q = $2.56~10^{53}$ photons .${\rm s}^{-1}$.}
\label{figure7}
\end{figure}

\begin{figure}
\begin{center}
\resizebox{\hsize}{14cm}{\includegraphics{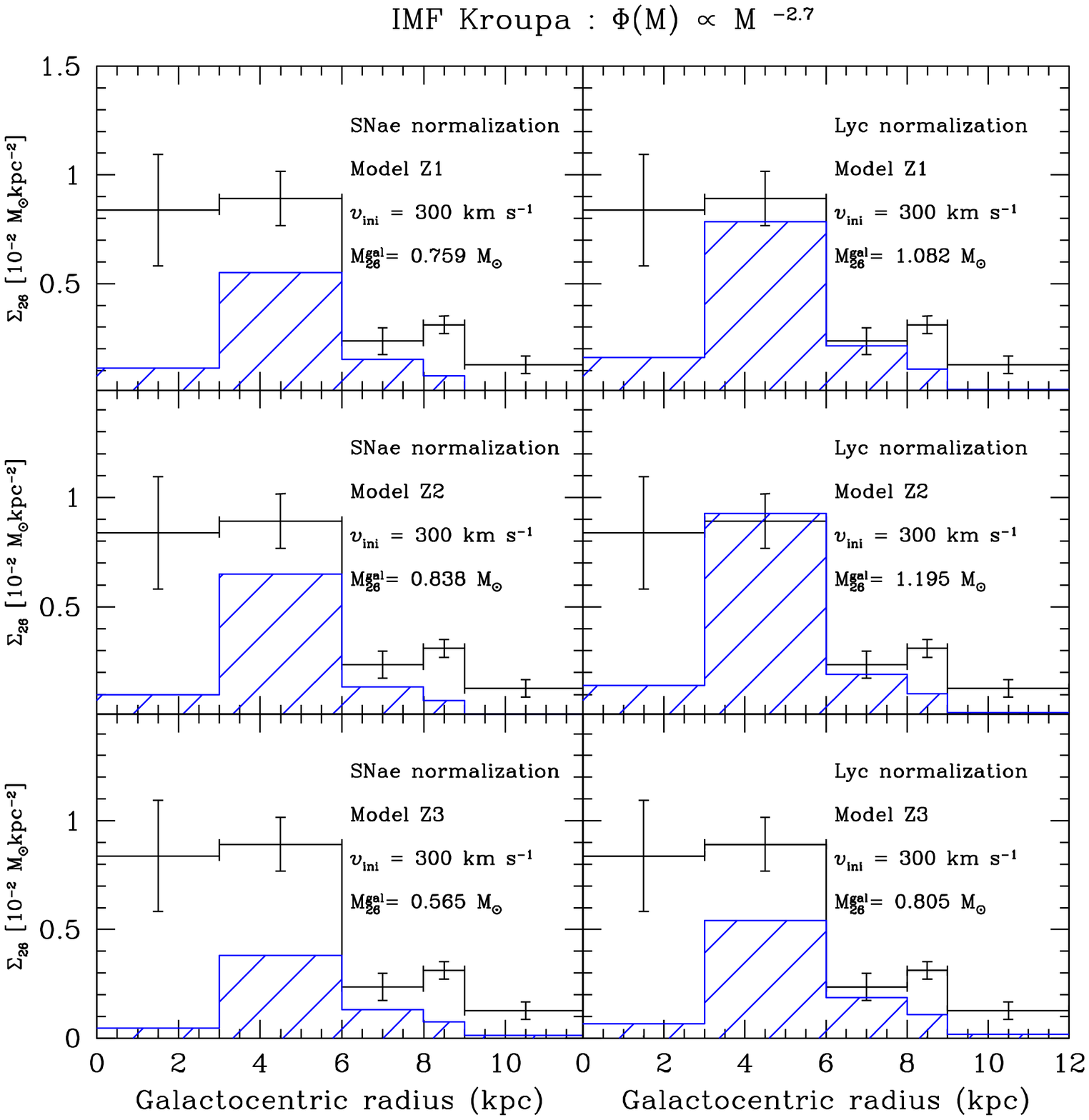}}
\end{center}
\caption{Same as right column of Fig.~6 and Fig.~7 for a Kroupa IMF.}
\label{figure8}
\end{figure}

 In Fig.~\ref{figure5}, Fig.~\ref{figure6}, Fig.~\ref{figure7} and
 Fig.~\ref{figure8}, we compare the theoretical mass surface density
 profiles of \al, $\Sigma_{26}$ (dashed histograms) with the profile
 deduced from COMPTEL data analysis (Kn\"odlseder \cite{Kn97}; black
 curve). In Fig.~\ref{figure5}, we use the SFR from Wang \& Silk
 (\cite{WS94}), normalized to a rate of 2 SNae per century. In
 Fig.~\ref{figure6} and Fig.~\ref{figure7} , the SFR is estimated following
 Fig.~3 of Paladini et al. (\cite{P04}), and it is normalized either on the
 SNae rate (Fig.~\ref{figure6}), or on the total galactic Lyman continuum
 luminosity (Fig.~\ref{figure7}). For each case we compare the effect of
 rotation and of the adopted galactic metallicity gradient. In Figures 5 to
 7, we used a Salpeter IMF; Fig.~\ref{figure8} is analogous to
 Fig.~\ref{figure6} (right panels) and Fig.~\ref{figure7} (right panels)
 but for a Kroupa IMF ($\Phi (M) \propto M^{-2.7}$).

 \begin{itemize}
 
 \item The theoretical predictions are in general below the observed
 surface density curve. This is expected since in our estimate we only
 account for the contribution of the winds of WR stars, while other sources
 as the supernovae likely also contribute.

  \item In the present theoretical framework, the fraction contributed by
 other sources could be estimated from the discrepancy between the
 theoretical and observed histograms. From Fig.~\ref{figure5} to
 Fig.~\ref{figure7}, we see that the contribution of other sources than WR
 stars is much larger when the yields of non--rotating models are accounted
 for. As recalled in Sect.~1 (see paper XI for more details), the
 non--rotating models fail to fit many observed properties of WR star
 populations that rotating models succeed in reproducing.  We will thus
 preferentially trust the results obtained from the rotating models.
 
 \item The higher the metallicity of the inner galactic regions (r $< 6$),
 the more important the WR contribution to the \al synthesis. In
 particular, models labelled as $Z3$, with the shallow metallicity
 gradient and an adopted value of 0.03 for the metallicity in the innermost
 regions, lead to the lower \al injection rates in all the cases presented
 here.
 
 \item In the outer regions, beyond 8 kpc, there is a deficit in all the
 models considered here when the initial rotation velocity assumed is that
 of $\upsilon_{\rm ini}=$ 300 km s$^{-1}$ .  If the initial distribution of
 rotation favours faster rotation at lower metallicity, such a deficit
 could be smaller, but at the present time the observational evidence for
 such a behaviour remains indirect (Maeder et al. \cite{MG99}).
 
 \item Around 50\% of the total mass of \al in the Galaxy originates from
 the regions between 3 kpc and 6 kpc, whatever the metallicity gradient
 adopted. This is related to the shape of the SFR, which presents a peak in
 this region (see Fig.~3 in Prantzos \& Aubert \cite{PA95} and in Paladini
 et al. \cite{P04}), mainly due to the presence of a ring of molecular
 clouds around 5 kpc.

\begin{table}
\caption{\al injection rates predicted by the different models computed :
  ({\bf A}) Salpeter's IMF, SFR from Wang \& Silk (1994), SNae
  normalization (Fig.~5); ({\bf B})
  Salpeter's IMF, SFR from Paladini et al. (2004), SNae normalization
  (Fig.~6); ({\bf C})
  Salpeter's IMF, SFR from Paladini et al. (2004), Lyc normalization
  (Fig.~7); ({\bf D})
  Kroupa's IMF, SFR from Paladini et al. (2004), Lyc normalization (Fig.~8); ({\bf E})
  Kroupa's IMF, SFR from Paladini et al. (2004), SNae normalization
  (Fig.~8).}
\begin{center}
\begin{tabular}{c|c|c|c|c|c|c}
\hline
\hline
{\scriptsize Model} & {\scriptsize $Z1$} & {\scriptsize $Z1$} & {\scriptsize
      $Z2$} & {\scriptsize $Z2$} & {\scriptsize $Z3$} & {\scriptsize $Z3$} \\\hline 
{\scriptsize $\upsilon_{ini}$} & {\scriptsize 0} & {\scriptsize 300} & {\scriptsize 0} & {\scriptsize 300} & {\scriptsize 0} & {\scriptsize 300} \\
 {\scriptsize (\kms)} & & & & & &\\\hline
{\scriptsize $\dot{M}_{26}$ ({\bf A})} &  {\scriptsize 0.563} &  {\scriptsize 1.242} &  {\scriptsize 0.617} &  {\scriptsize 1.336} &  {\scriptsize 0.346}
&  {\scriptsize 0.876} \\
{\scriptsize (\msun.${\rm Myr}^{-1}$)} & & & & & &\\\hline 
{\scriptsize $\dot{M}_{26}$ ({\bf B})} & {\scriptsize 0.551} & {\scriptsize 1.263} & {\scriptsize 0.632} & {\scriptsize 1.393} & {\scriptsize 0.371}
& {\scriptsize 0.943} \\
{\scriptsize (\msun.${\rm Myr}^{-1}$)} & & & & & &\\\hline 
{\scriptsize $\dot{M}_{26}$ ({\bf C})} & {\scriptsize 0.493} & {\scriptsize 1.130} & {\scriptsize 0.566} & {\scriptsize 1.247} & {\scriptsize 0.333}
& {\scriptsize 0.844}\\
{\scriptsize (\msun.${\rm Myr}^{-1}$)} & & & & & &\\\hline 
{\scriptsize $\dot{M}_{26}$ ({\bf D})} & {\scriptsize 0.434} & {\scriptsize 1.082} & {\scriptsize 0.500} & {\scriptsize 1.195} & {\scriptsize 0.292}
& {\scriptsize 0.805}\\
{\scriptsize (\msun.${\rm Myr}^{-1}$)} & & & & & &\\\hline 
{\scriptsize $\dot{M}_{26}$ ({\bf E})} & {\scriptsize 0.305} & {\scriptsize 0.759} & {\scriptsize 0.351} & {\scriptsize 0.838} & {\scriptsize 0.205}
& {\scriptsize 0.565}\\
{\scriptsize (\msun.${\rm Myr}^{-1}$)} & & & & & &\\\hline 
\end{tabular}
\end{center}
\end{table}

 \item The important dependence of the theoretical results on the adopted
 metallicity distribution in the Galaxy appears clearly through these
 numerical experiments. Depending on the metallicity distribution law
 adopted, both the radial distribution of the
 \al and the total quantity of \al predicted to be deposited by the WR stellar winds vary
 significantly. Shallower metallicity gradients, as adopted in model $Z3$,
 produce smaller values for M$_{26}^{\rm gal}$ and smoother profiles of
 $\Sigma_{26}$. This conclusion holds whether rotation is included or
 not.\\ Thus from the observed distribution of $\Sigma_{26}$, it is not
 possible to infer constraints on the metallicity dependence of the stellar
 yields (even if only one source dominates), unless more reliable and
 concordant estimates of the galactic metallicity gradients are obtained.

 \item Comparing Fig.~\ref{figure5}, Fig.~\ref{figure6} and
 Fig.~\ref{figure7} also allows us to study the effects of the choice of
 the SFR or that of its normalization on the \al distribution and injection
 rate by WR stars.  The SFR adopted entirely determines the shape of the
 surface density distribution predicted : while the Wang \& Silk
 (\cite{WS94}) prescription leads to a monotonous decrease of $\Sigma_{26}$
 with increasing galactocentric radius, the values derived from Paladini et
 al. (\cite{P04}) all present a low surface density in the inner regions, a
 peak in the region of the molecular clouds ring, and then a monotonous
 decrease, in better agreement with the nominal observed
 distribution\footnote{Note that within the errorbars from observations,
 the shape obtained with the Wang \& Silk (\cite{WS94}) prescription also
 matches the observed distribution.}.  On the other hand, as can be seen
 from Table 3, these differences between the two prescriptions used for the
 SFR (Fig.~\ref{figure5} and Fig.~\ref{figure6}) only slightly affect the
 total contributed mass of \al predicted, both for rotating and
 non--rotating WR models. \\ Changing the normalization, and adjusting the
 SFR to the match the observed galactic Lyman continuum luminosity does not
 change the results significantly, despite the fact that the value adopted
 of 2.56 photons .${\rm s}^{-1}$ corresponds to a lower rate of SNae
 events, namely 1.2 SNae per century. The masses deduced with the same
 value for Q as Kn\"odlseder (1999), i.e. 3.5 photons .${\rm s}^{-1}$ are
 to be multiplied by 1.37, which leads in the most favorable case to 1.70
 \msun of the present galactic mass of \al originating from WR stars.\\
 Finally, using an IMF slope of x = 1.7 (Scalo, 1986; Kroupa 1993) rather
 than 1.35 lowers the total galactic mass of \al predicted by the models as
 was shown in Fig.~\ref{figure8} and in Palacios et al. (\cite{PMV04} in
 their Fig.~3), without affecting the shape of the \al surface density
 distribution.\\

 \end{itemize}

 From the above estimates and comparisons, we can conclude that when
 stellar models accounting for the observed properties of the WR star
 populations are used ({\it i.e.} rotating models), WR stars appear to be
 significant if not dominant contributors to the present day \al content of
 the Milky Way.  The predictions for the total mass of \al originating from
 WR stars vary from 0.6 \msun to 1.4 \msun (see Table~3), whereas for the
 black histogram to which we compare our predictions, Kn\"odlseder (1997)
 obtained a total galactic mass of \al of 1.8 $\pm$ 0.1 \msun. The values
 listed in Table 3 show that WR stars can account for 20\% to 93\% of the
 observational estimates of the actual total galactic \al mass (1.5 - 3
 \msun). The predicted masses are little affected ($\approx$ 10\%
 variation) by the uncertainties on the galactic SFR, the IMF slope or the
 quantity against which the number of massive stars is to be normalized. On
 the other hand, the predicted total mass of \al originating from WR stars
 varies by as much as 30\% according to the galactic metallicity gradient
 adopted.  The shape of the predicted surface density distributions of \al,
 $\Sigma_{26}$, against the galactocentric radius strongly depends on the
 SFR adopted, so that the comparison with the values deduced from
 observations is of little interest.
 
 \section{Conclusion}

 We have studied the impact of rotation and initial metallicity on the
 yields of \al ejected by the WR stellar winds. When making use of rotation
 velocities corresponding to the observed average rotational velocities
 during the MS phase, {\it i.e.} for values around 200 \kms, the inclusion
 of rotation, all other things being equal, roughly doubles the WR
 contribution. Higher initial rotation velocities would still increase the
 quantity of \al ejected by the WR stellar winds.

 We compared the effects of rotation and those induced by an enhancement of
 the mass loss rates. On one hand, non--rotating models with enhanced mass
 loss rates (Meynet et al. \cite{Me97}) as well as the rotating models
 presented here (for which the mass loss rates are quite small), are both
 able to reproduce the observed variation with the metallicity of the
 number of WR to O--type stars. On the other hand, the \al yields predicted
 are different, especially in the lower mass domain. This is evidence of
 the fact that rotation has other more subtle effects rather than an
 enhancement of the mass loss rates, as already pointed out in Sect.~3. Our
 rotating models thus predict wind ejected masses that are about a factor
 of 1.4 superior to the ones provided by Meynet et al. (1997).

 We showed that the use of metallicity dependent mass loss rates during the
 WR phase as suggested by Crowther et al. (2000) does not have a
 significant impact on the results.

 The global contribution of WR stars remains difficult to assess in view of
 the numerous uncertainties pertaining not only to some physical
 ingredients of the models as the mass loss rates or the nuclear reaction
 rates, but also to galactic parameters such as the galactic metallicity
 gradient or the supernova rate.  We showed that the choice of the SFR
 tracer determines the shape of the \al surface density distribution with
 galactocentric distance, but marginally affects the predicted contributed
 mass of \al by WR. On the other hand, the choice of the metallicity
 gradient as well as that of the IMF slope may significantly affect the
 contribution of WR stars to the total galactic mass of \al, which can
 amount to 0.6 \msun (shallow metallicity gradient with Kroupa's IMF) up to
 1.4 \msun (steep metallicity gradient with Salpeter's IMF).

 In that respect the study of associations such as the Cygnus region,
 sufficiently young to prevent supernovae having much contributed to the
 \al enrichment are of prime importance. Possibly the detection of the
 1.809 MeV emission around a single source would be even more important.
 In that respect the upper limit found for $\gamma^2$--Vel (Oberlack et
 al. \cite{O00}) already provides a strong constraint. The binary nature of
 the object however requires some care when comparisons are made with a
 single star model. This particular case will be studied in detail in a
 forthcoming paper.

 \acknowledgements
 A. Palacios acknowledges financial support from ESA PRODEX fellowship
 ${\rm n}^{\rm o}$ 90069.

\end{document}